\documentclass[12pt]{article}
\usepackage{epsfig}
\def\be{\begin{equation}}
\def\ee{\end{equation}}
\def\bea{\begin{eqnarray}}
\def\eea{\end{eqnarray}}
\usepackage{graphicx}% Include figure files

\catcode`\@=11
\def\lsim{\mathrel{\mathpalette\@versim<}}
\def\gsim{\mathrel{\mathpalette\@versim>}}
\def\@versim#1#2{\vcenter{\offinterlineskip
\ialign{$\m@th#1\hfil##\hfil$\crcr#2\crcr\sim\crcr } }}
\catcode`\@=12

\catcode`@=12
\begin{document}
\thispagestyle{empty}
\begin{flushright}
UCRHEP-T551\\
February 2015\
\end{flushright}
\vspace{0.6in}
\begin{center}
{\LARGE \bf Neutrino Mixing and CP Phase Correlations\\}
\vspace{1.5in}
{\bf Ernest Ma, Alexander Natale, and Oleg Popov\\}
\vspace{0.1in}
{\sl Department of Physics and Astronomy,\\}
\vspace{0.1in}
{\sl University of California, 
Riverside, California 92521, USA\\}
\end{center}
\vspace{1.5in}

\begin{abstract}\
A special form of the $3 \times 3$ Majorana neutrino mass matrix derivable 
from $\mu - \tau$ interchange symmetry accompanied by a generalized $CP$ 
transformation was obtained many years ago.  It predicts $\theta_{23} = \pi/4$ 
as well as $\delta_{CP} = \pm \pi/2$, with $\theta_{13} \neq 0$.  Whereas 
this is consistent with present data, we explore a deviation of this 
result which occurs naturally in a recent proposed model of radiative 
inverse seesaw neutrino mass.
\end{abstract}

\newpage
\baselineskip 24pt
A special form of the $3 \times 3$ Majorana neutrino mass matrix first 
appeared in 2002~\cite{m02,bmv03}, i.e. 
\begin{equation}
{\cal M}_\nu = \pmatrix{A & C & C^* \cr C & D^* & B \cr C^* & B & D},
\end{equation}
where $A,B$ are real.  It was shown that $\theta_{13} \neq 0$ and yet 
both $\theta_{23}$ and the $CP$ nonconserving phase $\delta_{CP}$ are 
maximal, i.e. $\theta_{23} = \pi/4$ and $\delta_{CP} = \pm \pi/2$. 
Subsequently, this pattern was shown~\cite{gl04} to be protected by a 
symmetry, i.e. $e \to e$ and $\mu \leftrightarrow \tau$ exchange with 
$CP$ conjugation.  All three predictions are consistent with present 
experimental data.  Recently, a radiative (scotogenic) model of inverse 
seesaw neutrino mass has been proposed~\cite{fmp14} which naturally obtains 
\begin{equation}
{\cal M}_\nu^\lambda = \pmatrix{1 & 0 & 0 \cr 0 & 1 & 0 \cr 0 & 0 & \lambda} 
{\cal M}_\nu \pmatrix{1 & 0 & 0 \cr 0 & 1 & 0 \cr 0 & 0 & \lambda},
\end{equation}
where $\lambda = f_\tau/f_\mu$ is the ratio of two real Yukawa couplings.

This model has three real singlet scalars $s_{1,2,3}$ and one Dirac fermion 
doublet $(E^0,E^-)$ and one Dirac fermion singlet $N$, all of which are 
odd under an exactly conserved (dark) $Z_2$ symmetry.  As a result, the 
third one-loop radiative mechanism proposed in 1998~\cite{m98} for 
generating neutrino mass is realized, as shown below.
\begin{figure}[htb]
\vspace*{-3cm}
\hspace*{-3cm}
\includegraphics[scale=1.0]{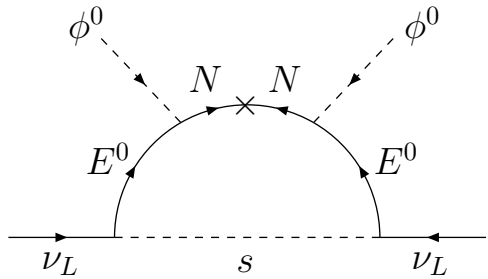}
\vspace*{-21.5cm}
\caption{One-loop generation of inverse seesaw neutrino mass.}
\end{figure}

The mass matrix linking $(\bar{N}_L, \bar{E}^0_L)$ to $(N_R, E^0_R)$ 
is given by
\begin{equation}
{\cal M}_{N,E} = \pmatrix{m_N & m_D \cr m_F & m_E},
\end{equation}
where $m_N,m_E$ are invariant mass terms, and $m_D,m_F$ come from the Higgs 
vacuum expectation value $\langle \phi^0 \rangle = v/\sqrt{2}$.  As a result, 
$N$ and $E^0$ mix to form two Dirac fermions of masses $m_{1,2}$, with 
mixing angles
\begin{eqnarray}
m_D m_E + m_F m_N &=& \sin \theta_L \cos \theta_L (m_1^2 - m_2^2), \\
m_D m_N + m_F m_E &=& \sin \theta_R \cos \theta_R (m_1^2 - m_2^2).
\end{eqnarray}
To connect the loop, Majorana mass terms $(m_L/2) N_L N_L$ and 
$(m_R/2) N_R N_R$ are assumed.  Since both $E$ and $N$ may be defined 
to carry lepton number, these new terms violate lepton number softly and 
may be naturally small, thus realizing the mechanism of inverse 
seesaw~\cite{ww83,mv86,m87} as explained in Ref.~\cite{fmp14}.  
Using the Yukawa interaction $f s \bar{E}^0_R \nu_L$, the one-loop 
Majorana neutrino mass is given by 
\begin{eqnarray}
m_\nu &=& f^2 m_R \sin^2 \theta_R \cos^2 \theta_R (m_1^2 - m_2^2)^2 
\int {d^4 k \over (2 \pi)^4} {k^2 \over (k^2 - m_s^2)} {1 \over 
(k^2 - m_1^2)^2} {1 \over (k^2 - m_2^2)^2} \nonumber \\ 
&+& f^2 m_L m_1^2 \sin^2 \theta_L \cos^2 \theta_R \int {d^4 k \over (2 \pi)^4} 
{1 \over (k^2 - m_s^2)}{1 \over (k^2 - m_1^2)^2} \nonumber \\ 
&+& f^2 m_L m_2^2 \sin^2 \theta_R \cos^2 \theta_L \int {d^4 k \over (2 \pi)^4}
{1 \over (k^2 - m_s^2)}{1 \over (k^2 - m_2^2)^2} \nonumber \\ 
&-& 2 f^2 m_L m_1 m_2 \sin \theta_L \sin \theta_R \cos \theta_L \cos \theta_R 
\int {d^4 k \over (2 \pi)^4}{1 \over (k^2 - m_s^2)}{1 \over (k^2 - m_1^2)}
{1 \over (k^2 - m_2^2)}.   
\end{eqnarray}
It was also shown in Ref.~\cite{fmp14} that the 
implementation of a discrete flavor $Z_3$ symmetry, which is softly 
broken by the $3 \times 3$ real scalar mass matrix spanning $s_{1,2,3}$, 
leads to ${\cal M}_\nu^\lambda$ of Eq.~(2).

To explore how the predictions $\theta_{23} = \pi/4$ and $\delta_{CP} = 
\pm \pi/2$ are changed for $\lambda \neq 1$, consider the general 
diagonalization of ${\cal M}_\nu$, i.e. 
\begin{equation}
{\cal M}_\nu = E_\alpha U E_\beta {\cal M}_d E_\beta U^T E_\alpha,
\end{equation}
where
\begin{equation}
E_\alpha = \pmatrix{e^{i \alpha_1} & 0 & 0 \cr 0 & e^{i \alpha_2} & 0 \cr 
0 & 0 & e^{i \alpha_3}}, ~~~ 
E_\beta = \pmatrix{e^{i \beta_1} & 0 & 0 \cr 0 & e^{i \beta_2} & 0 \cr 
0 & 0 & e^{i \beta_3}}, ~~~ {\cal M}_d = \pmatrix{m_1 & 0 & 0 \cr 
0 & m_2 & 0 \cr 0 & 0 & m_3}. 
\end{equation}
Hence
\begin{equation}
{\cal M}_\nu {\cal M}_\nu^\dagger = E_\alpha U {\cal M}_d^2 U^\dagger 
E_\alpha^\dagger.
\end{equation}
We then have
\begin{equation}
{\cal M}_\nu^\lambda ({\cal M}_\nu^\lambda)^\dagger  = E_\alpha U [1 + \Delta] 
{\cal M}_{\lambda d}^2 [1 + \Delta^\dagger] U^\dagger E_\alpha^\dagger,
\end{equation}
where
\begin{equation}
\Delta = U^\dagger \pmatrix{0 & 0 & 0 \cr 0 & 0 & 0 \cr 0 & 0 &  \lambda-1} 
U, ~~~ {\cal M}_{\lambda d}^2 = 
\pmatrix{m_1^2 & 0 & 0 \cr 0 & m_2^2 & 0 \cr 0 & 0 & \lambda^2 m_3^2}.
\end{equation}
We now diagonalize numerically
\begin{equation}
[1 + \Delta] {\cal M}^2_{\lambda d} [ 1 + \Delta^\dagger] = 
O {\cal M}^2_{new} O^T,
\end{equation}
where $O$ is an orthogonal matrix, and ${\cal M}^2_{new}$ is diagonal with mass 
eigenvalues equal to the squares of the physical neutrino masses.  
Let us define
\begin{equation}
A = (1 + \Delta)^{-1} O,
\end{equation}
then
\begin{equation}
A {\cal M}^2_{new} A^\dagger = {\cal M}^2_{\lambda d}.
\end{equation}
Since $U$ is known with $\theta_{23} = \pi/4$ and $\delta = \pm \pi/2$, 
we know $\Delta$ once $\lambda$ is chosen.  The orthogonal matrix $O$ 
has three angles as parameters, so $A$ has three parameters.  In Eq.~(14), 
once the three physical neutrino mass eigenvalues of ${\cal M}^2_{new}$ are 
given, the three off-diagonal entries of ${\cal M}^2_{\lambda d}$ are 
constrained to be zero, thus determining the three unknown parameters 
of $O$.  Once $O$ is known, $U O$ is the new neutrino mixing matrix,
from which we can extract the correlation of $\theta_{23}$ with $\delta_{CP}$.
There is of course an ambiguity in choosing the three physical neutrino 
masses, since only $\Delta m^2_{32}$ and $\Delta m^2_{21}$ are known.  
There are also the two different choices of $m_1 < m_2 < m_3$ (normal ordering) 
and $m_3 < m_1 < m_2$ (inverted ordering).  We consider each case, and 
choose a value of either $m_1$ or $m_3$ starting from zero.  We then obtain 
numerically the values of $\sin^2 (2 \theta_{23})$ and $\delta_{CP}$ as 
functions of $\lambda \neq 1$.  We need also to adjust  
the input values of $\theta_{12}$ and $\theta_{13}$, so that their output 
values for $\lambda \neq 1$ are the preferred experimental values.

\begin{figure}[htb]
%\vspace*{1cm}
\hspace*{1cm}
\includegraphics[scale=1.25]{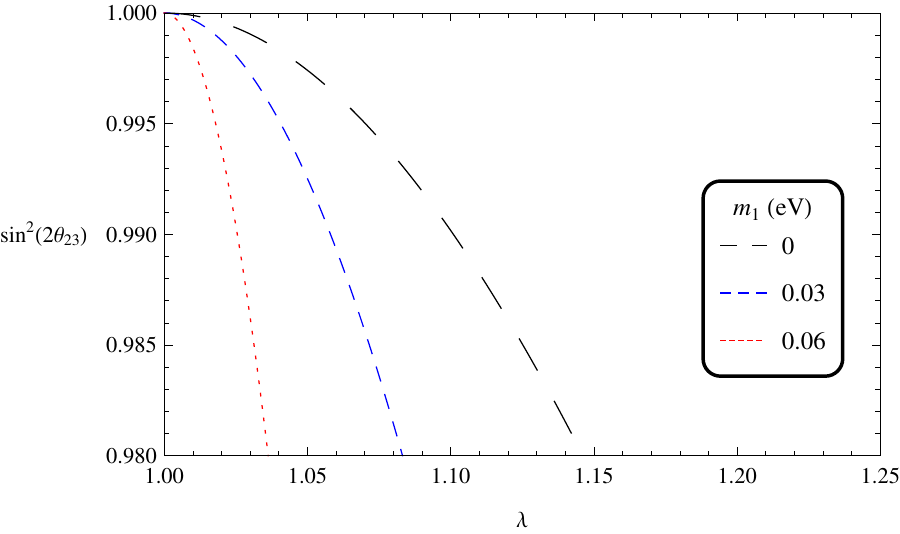}
%\vspace*{-21.5cm}
\caption{$\sin^2 (2 \theta_{23})$ versus $\lambda$ in normal ordering.}
\end{figure}
\begin{figure}[htb]
%\vspace*{-3cm}
\hspace*{2cm}
\includegraphics[scale=1.15]{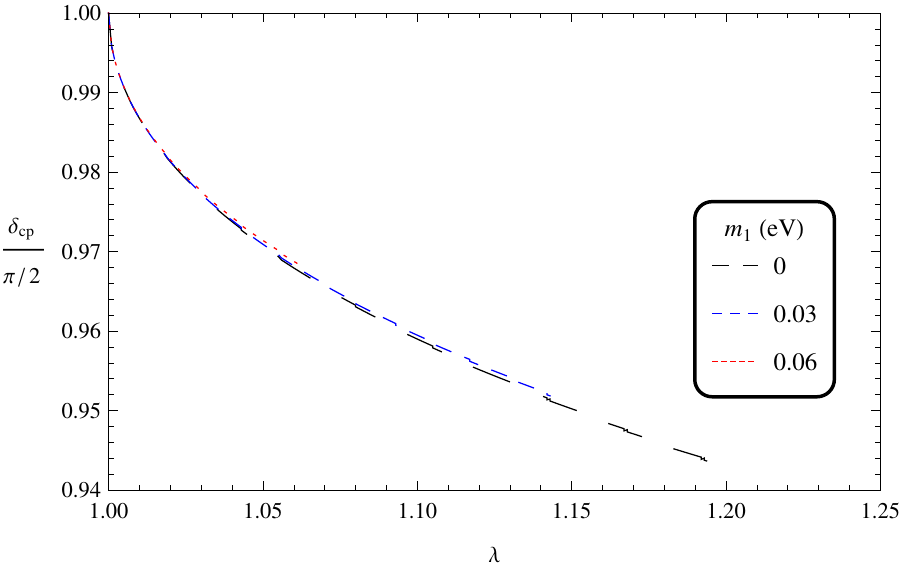}
%\vspace*{-21.5cm}
\caption{$\delta_{CP}$ versus $\lambda$ in normal ordering.}
\end{figure}

We use the 2014 Particle Data Group values~\cite{pdg2014} of neutrino 
parameters:
\begin{eqnarray}
&& \sin^2 (2 \theta_{12}) = 0.846 \pm 0.021, ~ \Delta m^2_{21} = (7.53 \pm 
0.18) \times 10^{-5}~{\rm eV}^2, \\ 
&& \sin^2 (2 \theta_{23}) = 0.999 \pmatrix{+0.001 \cr -0.018}, ~  
\Delta m^2_{32} = (2.44 \pm 0.06) \times 10^{-3}~{\rm eV}^2 ~({\rm normal}), \\ 
&& \sin^2 (2 \theta_{23}) = 1.000 \pmatrix{+0.000 \cr -0.017}, ~  
\Delta m^2_{32} = (2.52 \pm 0.07) \times 10^{-3}~{\rm eV}^2 ~({\rm inverted}), \\
&& \sin^2 (2 \theta_{13}) = (9.3 \pm 0.8) \times 10^{-2}.
\end{eqnarray}

We consider first normal ordering, choosing the three representative values 
$m_1 = 0,0.03,0.06$ eV.  We then vary the value of $\lambda > 1$.  [The case 
$\lambda < 1$ is equivalent to $\lambda^{-1} > 1$ with $\mu -\tau$ exchange.] 
Following the algorithm already mentioned, we obtain numerically the values 
of $\sin^2 (2 \theta_{23})$ and $\delta_{CP}$ as functions of $\lambda$.  Our 
solutions are fixed by the central values of $\Delta m^2_{21}$, 
$\Delta m^2_{32}$, $\sin^2 (2 \theta_{12})$, and $\sin^2 (2 \theta_{13})$. 
In Figs.~2 and 3 we plot $\sin^2 (2 \theta_{23})$ and $\delta_{CP}$ respectively 
versus $\lambda$.  We see from Fig.~2 that $\lambda < 1.15$ is required for 
$\sin^2 (2 \theta_{23}) > 0.98$.  We also see from Fig.~3 that $\delta_{CP}$ is 
not sensitive to $m_1$.  Note that our scheme does not distinguish 
$\delta_{CP}$ from $-\delta_{CP}$.  In Fig.~4 we plot $\sin^2 (2 \theta_{23})$ 
versus $\delta_{CP}$.  We see that $\delta_{CP}/(\pi/2) > 0.95$ is required 
for $\sin^2 (2 \theta_{23}) > 0.98$.
\begin{figure}[htb]
%\vspace*{-3cm}
\hspace*{1cm}
\includegraphics[scale=1.3]{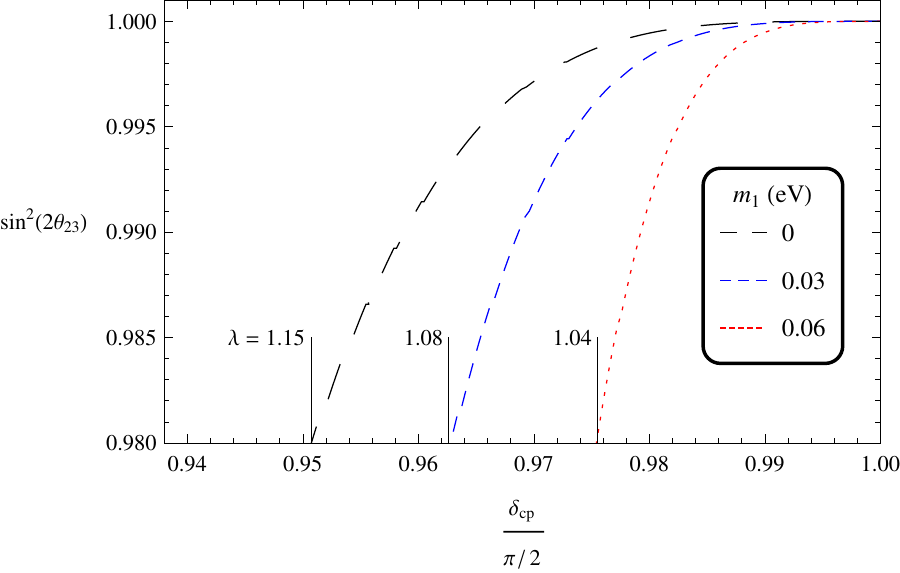}
%\vspace*{-21.5cm}
\caption{$\sin^2 (2 \theta_{23})$ versus $\delta_{CP}$ in normal ordering.}
\end{figure}

We then consider inverted ordering, using $m_3$ instead of $m_1$.  We plot 
in Figs.~5, 6, and 7 the corresponding results.  Note that 
in our scheme, the effective neutrino mass $m_{ee}$ measured in neutrinoless 
double beta decay is very close to $m_1$ in normal ordering and $m_3 + 
\sqrt{\Delta m^2_{32}}$ in inverted ordering.  
We see similar constraints on $\sin^2 (2 \theta_{23})$ and $\delta_{CP}$. 
In other words, our scheme is insensitive to whether normal or inverted 
ordering is chosen.  Finally, we have checked numerically that 
$\theta_{23} < \pi/4$ if $\lambda > 1$, and $\theta_{23} > \pi/4$ if 
$\lambda < 1$.  As we already mentioned, the two solutions are related by 
the mapping $\lambda \to \lambda^{-1}$. 

\begin{figure}[htb]
%\vspace*{-3cm}
\hspace*{1cm}
\includegraphics[scale=1.25]{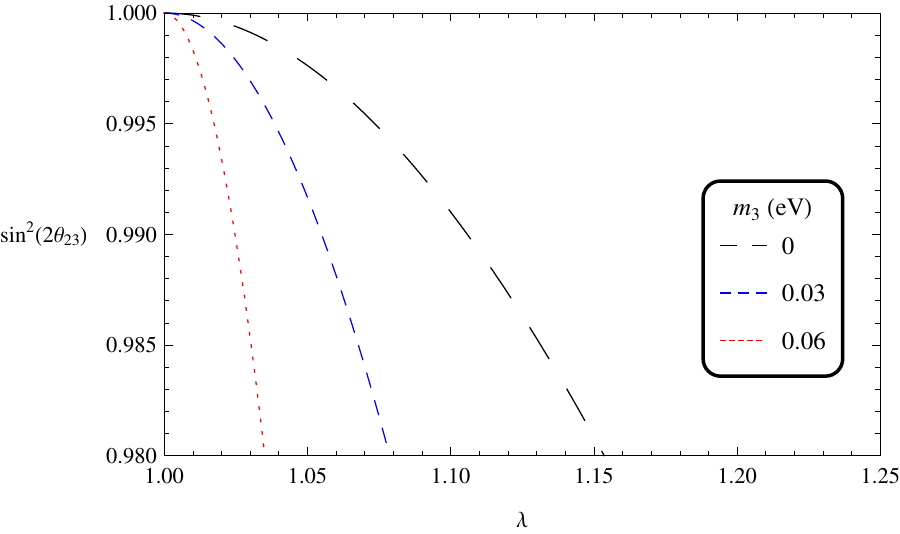}
%\vspace*{-21.5cm}
\caption{$\sin^2 (2 \theta_{23})$ versus $\lambda$ in inverted ordering.}
\end{figure}
\begin{figure}[htb]
%\vspace*{-3cm}
\hspace*{2cm}
\includegraphics[scale=1.15]{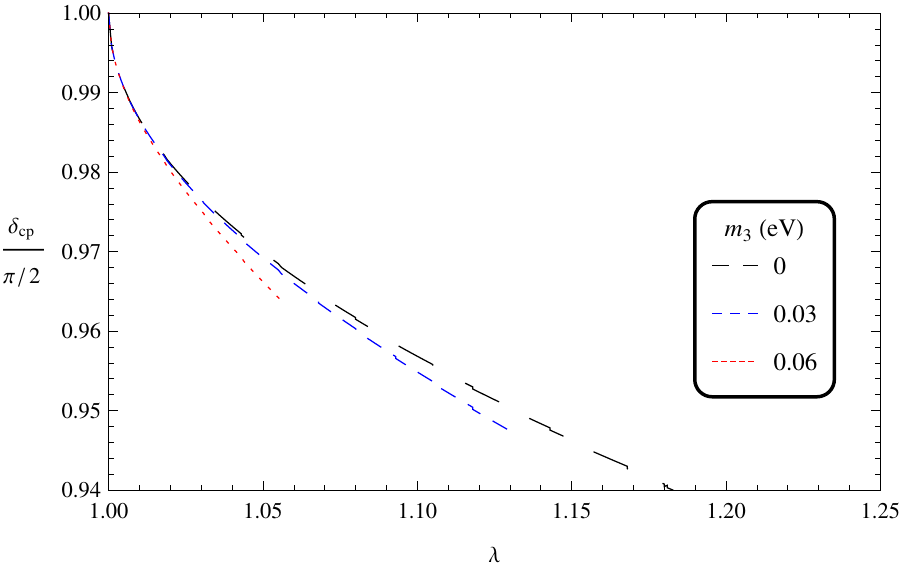}
%\vspace*{-21.5cm}
\caption{$\delta_{CP}$ versus $\lambda$ in inverted ordering.}
\end{figure}

\newpage
\begin{figure}[htb]
%\vspace*{-3cm}
\hspace*{1cm}
\includegraphics[scale=1.3]{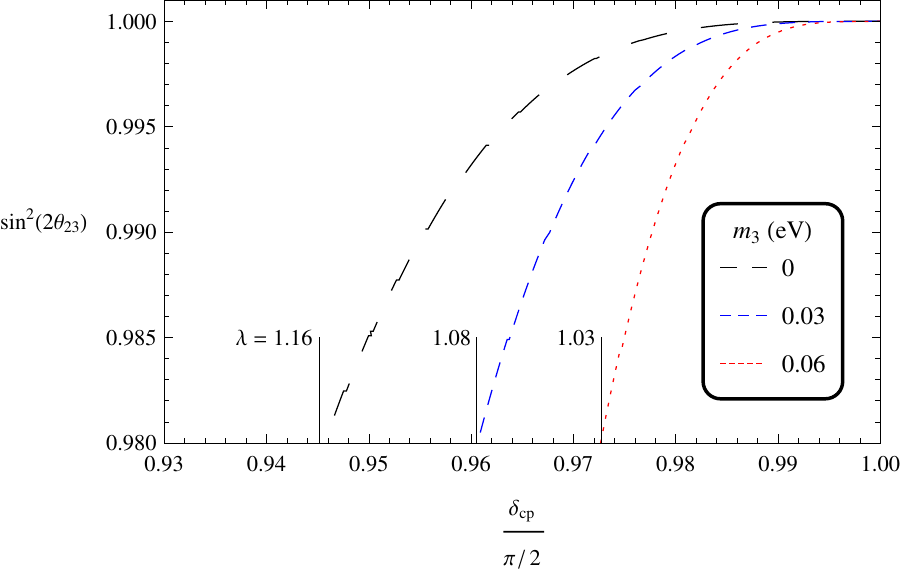}
%\vspace*{-21.5cm}
\caption{$\sin^2 (2 \theta_{23})$ versus $\delta_{CP}$ in inverted ordering.}
\end{figure}
In conclusion, we have explored the possible deviation from the prediction 
of maximal $\theta_{23}$ and maximal $\delta_{CP}$ in a model of radiative 
inverse seesaw neutrino mass.  We find that given the present $1 \sigma$ 
bound of $0.98$ on $\sin^2 (2 \theta_{23})$, $\delta_{CP}/(\pi/2)$ must be 
greater than about 0.95.

%\newpage
This work is supported in part 
by the U.~S.~Department of Energy under Grant No.~DE-SC0008541.

\bibliographystyle{unsrt}

\end{document}